\RequirePackage{lineno}
\documentclass[review,number,sort&compress,aps,showpacs,showkeys,amsmath,amssymb,superscriptaddress,groupedaddress]{revtex4}  
\usepackage{setspace}
\usepackage{graphicx}  
\usepackage{dcolumn}   
\usepackage{bm}        
\usepackage{amssymb}   
\usepackage{epstopdf}
\usepackage{amsmath}
\usepackage{cases}
\usepackage{lineno}
\usepackage{hyperref}

\hypersetup{colorlinks=true}
\newcommand{\D}{\displaystyle}
\newcommand{\DF}[2]{\frac{\D#1}{\D#2}}

\begin{document}
\begin{spacing}{2.0}
\title{Determination of the total absorption peak in an electromagnetic calorimeter}

\newcommand{\TsingHua}{\affiliation{Department~of~Engineering~Physics, Tsinghua~University, Beijing}}
\newcommand{\Key}{\affiliation{Key Laboratory of Particle \& Radiation Imaging (Tsinghua University), Ministry of Education}}
\newcommand{\NCTU}{\affiliation{Institute~of~Physics, National~Chiao-Tung~University, Hsinchu}}

\author{Jia-Hua~Cheng}\NCTU
\author{Zhe~Wang}\TsingHua\Key
\author{Logan~Lebanowski}\TsingHua\Key
\author{Guey-Lin~Lin}\NCTU
\author{Shaomin~Chen}\TsingHua\Key

\date{March 14, 2016}

\begin{abstract}
A physically-motivated function was developed to accurately determine the total absorption peak in an electromagnetic calorimeter and to overcome
biases present in many commonly used methods.
The function is the convolution of a detector resolution function with the sum of a delta function, which represents the complete absorption of energy, and a tail function, which describes the partial absorption of energy and depends on the detector materials and structures.
Its performance was tested with the simulation of three typical cases.
The accuracy of the extracted peak value, resolution, and peak area was improved by an order of magnitude on average,
relative to the Crystal Ball function. 
\end{abstract}


\maketitle

\section{Introduction}
The extraction of useful and physically meaningful parameters from a measured energy spectrum is a long-standing question in particle and nuclear physics experiments. Electromagnetic (EM) calorimeters have been a cornerstone of particle physics experiments and today are still irreplaceable as tools for making new discoveries, like at the Large Hadron Collider~\cite{LHC}.
Because of their finite sizes and the presence of insensitive materials, EM calorimeters do not faithfully convey the incident energy of a particle in many cases.
This fact can be significant for experiments that require relatively high accuracy in energy, like the RENO50~\cite{RENO50}, JUNO~\cite{JUNO}, and Jinping~\cite{Jinping} neutrino experiments.

Several functions are frequently used to fit the spectrum of deposited energy in electromagnetic calorimeters.
The Gaussian function and truncated Gaussian function are very popular in determining the total absorption peak.
Another widely used function originally introduced for this purpose is the Crystal Ball (CB) function~\cite{CB1, CB2, CB3}.
The logarithmic Gaussian and exponentially modified Gaussian distribution have also been used, for example in Refs.~\cite{book, EMG, Bukin}.
The total absorption peak extracted using these functions has intrinsic biases.

In this article, a detailed discussion is given with respect to the CB function without loss of generality.
The CB function was developed to describe the energy deposition of electrons or gammas in an array of NaI (Tl) crystals
of the calorimeter of the Crystal Ball experiment.
It has been used, for example, to describe EM energy deposition in the liquid scintillator detectors
of the Daya Bay neutrino experiment~\cite{DYB}
and the invariant mass of a resonance with radiative energy loss in the
Belle~\cite{Belle} and LHCb~\cite{LHCb} experiments.
It has become one of the standard functions supported by popular
analysis softwares, for example, RooFit of ROOT~\cite{RooFit}.

The CB function is piecewise-defined, consisting of a Gaussian peak and a power-law tail.
The function can be expressed as
\begin{equation}
\begin{aligned}
f_{\rm{CB}}(x; \mu, \sigma, \alpha, n) = \ & N \left\{
\begin{matrix}
e^{-\DF{(x-\mu) ^2}{2\sigma^{2}}},
& \text{for}\ \DF{x-\mu}{\sigma} > -\alpha
\\
A\left(\DF{n}{\left | \alpha \right |}-\left | \alpha \right |-\DF{x-\mu}{\sigma}\right)^{-n},
& \text{for}\ \DF{x-\mu}{\sigma} \leqslant -\alpha
\end{matrix}\right. \\
& \text{with}\ A=\left(\DF{n}{\left | \alpha \right |}\right)^n e^{-\DF{\left |\alpha  \right |^2}{2}},
\end{aligned}
\end{equation}
where $\mu$ and $\sigma$ are the mean and standard deviation of the Gaussian peak,
$n$ is the exponent of the tail function,
$\alpha$ gives the connecting point of the Gaussian and tail function,
and $N$ is a normalization factor.
$\mu$ and $\sigma$ are generally used to study the energy scale and resolution of a calorimeter.

With the high statistics often accumulated in experiments or simulations, we observed
a consistently poor $\chi^2$ per number of degrees of freedom (NDF) when fitting with the CB function, and significant biases of the extracted means and resolutions, which will be demonstrated.

Theoretically it is difficult to find a physical motivation for two piecewise-defined sub-functions.  
The impact of the tail to the fitted mean and standard deviation of the Gaussian is indirect and obscure.
Furthermore, there is no apparent representation of the number of totally absorbed events.

In this article, we briefly discuss 
the energy deposition of photons and electrons in an EM calorimeter and introduce a physically-motivated EM calorimeter function in section~\ref{sec:newfunc},
present performance studies using \textsc{Geant4}~\cite{Geant4} simulation for three typical applications of
EM calorimeters with photons up to 50 MeV in section~\ref{sec:performance},
discuss additional details of the calorimeter function in section~\ref{sec:discussion}, and conclude in section~\ref{sec:conclusion}.



\section{Electromagnetic Calorimeter function}
\label{sec:newfunc}
The energy measured in an electromagnetic calorimeter is described with the electromagnetic calorimeter function (calorimeter function for short) $f_{\rm{cal}}$:
\begin{equation}
\label{eq:original}
f_{\rm{cal}} \equiv f_{\rm{energy\ deposited}}\otimes f_{\rm{resolution}},
\end{equation}
which is a convolution of the deposited energy spectrum and the detector resolution.  Detector nonlinearity and nonuniformity are not directly considered.

\begin{figure}[!h]
	\begin{center}
		\includegraphics[width=0.47\columnwidth, clip]{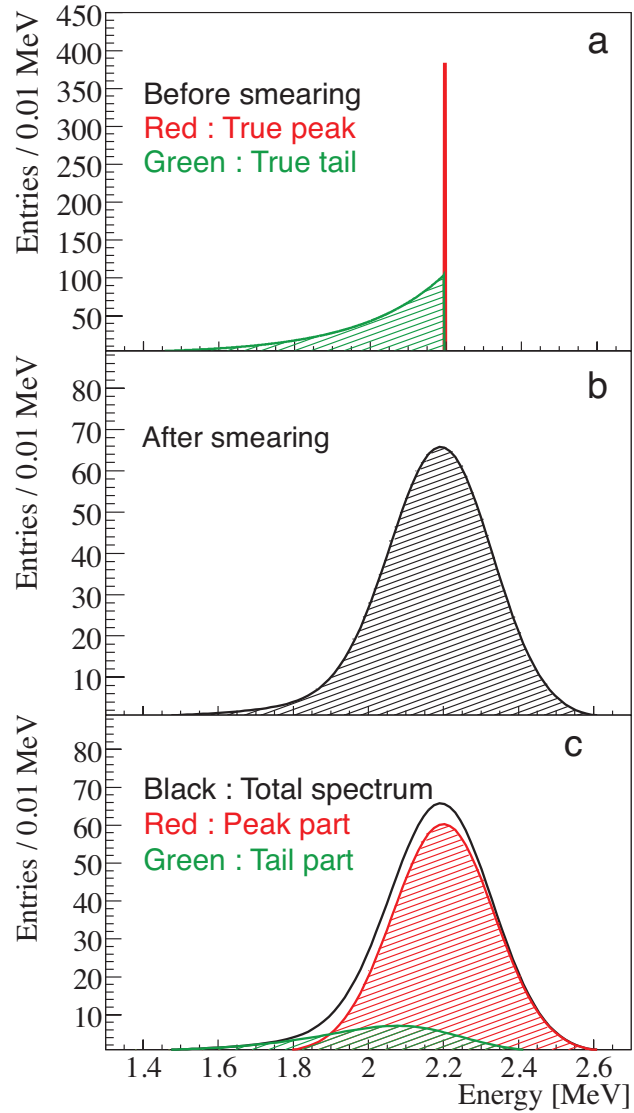}
	\end{center}
\caption{(Color online) (a) A typical distribution of the energy deposited by monoenergetic gammas or electrons in a calorimeter, including a
delta function for complete energy absorption and a tail for partial energy absorption.
(b) Measured spectrum obtained by applying Gaussian smearing to panel (a).
(c) A decomposition of the contributions of peak and tail of the measured spectrum.}
\label{fig:fig1group}
\end{figure}
Figure~\ref{fig:fig1group}(a) shows a typical distribution of the energy deposited by monoenergetic gammas or electrons in an EM calorimeter.
The distribution has some noticeable features:
\begin{itemize}
\item the peak is a Kronecker delta function corresponding to the complete absorption of the energy;
\item the tail to the left corresponds to the partial absorption of energy in the sensitive region of the detector;
\item the shape and fraction of the tail depends on the actual detector dimensions, materials, and structure.
\end{itemize}

Detector energy resolution can often be approximated by a Gaussian function and is generally a function of energy.
Figure~\ref{fig:fig1group}(b) gives the measured spectrum after a constant-resolution Gaussian smearing.

Figure~\ref{fig:fig1group}(c) illustrates the contributions of the peak and tail portions after the Gaussian smearing.
The smeared tail spreads into the Gaussian peak from the left, causing the maximum of the spectrum to occur below the
true Gaussian peak energy, and apparently enlarging the Gaussian resolution.  Fitting the spectrum without considering these effects
yields intrinsic biases.

Inspired by these observations, Eq.~(\ref{eq:original}) can take a more concrete form:
\begin{equation}
\begin{aligned}
\label{eq:derived}
f_{\rm{cal}} &= [\alpha f_{\rm{peak}}+(1-\alpha)f_{\rm{tail}}]\otimes f_{\rm{resolution}}\\
  &= [\alpha \delta +(1-\alpha)f_{\rm{tail}}]\otimes \text{Gauss},
\end{aligned}
\end{equation}
where $\alpha$ is the peak fraction and $f_{\rm{tail}}$ is the tail distribution, which is determined according to the situation.


\textbf{Peak:}
The sharp peak in Fig.~\ref{fig:fig1group}(a) corresponds to total energy absorption.
With a Gaussian detector resolution smearing, the shape of the peak is
\begin{equation}
\begin{aligned}
   \delta \otimes \text{Gauss} = \DF{1}{\sigma \sqrt{2\pi}} e^{-\frac{(E-\mu) ^2}{2\sigma^{2}} },
\label{eq:delta peak}
\end{aligned}
\end{equation}
where $E$ refers to the measured energy, and $\mu$ and $\sigma$ are the total absorption energy and energy resolution, respectively.

\textbf{Tail:}
The shape of the tail depends on the detector. Two analytical expressions are presented for demonstration:
Gaussian smearing of a constant function and Gaussian smearing of an exponential function.

For a constant function $C$, the measured energy distribution is
\begin{equation}
\begin{aligned}
   f_{\rm{const}} \otimes \text{Gauss}
   &= \int_0^{\mu} C\cdot\DF{1}{\sigma \sqrt{2\pi}} e^{-\frac{(E'-E) ^2}{2\sigma^{2}} }dE'\\
   &= \frac{C}{2}\left[\text{erf}\left(\frac{\mu-E}{\sqrt{2}\sigma}\right)-\text{erf}\left(\frac{-E}{\sqrt{2}\sigma}\right)\right],
\label{eq:const.tail}
\end{aligned}
\end{equation}
where $C$ is limited to within [0, $\mu$], and
erf is the Gaussian error function: $\mathrm{erf}(E) = \frac{2}{\sqrt\pi} \int_0^E \mathrm{exp}(-t^2) dt$.

For an exponential tail, the measured energy distribution is
\begin{equation}
\begin{aligned}
f _{\rm{exp}} \otimes \text{Gauss} \ &= \int_{0}^{\mu} \lambda e^{\lambda E'}\cdot\DF{1}{\sigma \sqrt{2\pi}}
        e^{-\frac{(E'-E) ^2}{2\sigma^{2}}}dE' \\
   &= \frac{\lambda}{2}e^\frac{\sigma^2\lambda^2+2\lambda E}{2}
        \left[
            \text{erf} \left( \frac{\mu-E-\sigma ^2\lambda}{\sqrt{2}\sigma} \right)-
            \text{erf} \left( \frac{-E-\sigma ^2\lambda}{\sqrt{2}\sigma} \right)
        \right],
\label{eq:exp.tail}
\end{aligned}
\end{equation}
where $\lambda$ is the slope of the exponential.

\textbf{Complete function:}
Combining Eqs.~(\ref{eq:derived}) and (\ref{eq:delta peak}) with Eq.~(\ref{eq:const.tail}) or Eq.~(\ref{eq:exp.tail}), we get
Eq.~(\ref{eq:const}) or
Eq.~(\ref{eq:exp}) for the complete calorimeter function with a constant or exponential tail, respectively.

\begin{widetext}
\begin{equation}
\begin{aligned}
f&_{\text{cal}}(E; \mu, \sigma, \alpha)
  =
   \alpha\DF{1}{\sigma \sqrt{2\pi}} e^{-\frac{(E-\mu) ^2}{2\sigma^{2}} }
  +(1-\alpha) \frac{1}{\mu}\left[\text{erf}\left(\frac{\mu-E}{\sqrt{2}\sigma}\right)-\text{erf}\left(\frac{-E}{\sqrt{2}\sigma}\right)\right] \rm{\ (constant\ tail),}
\label{eq:const}
\end{aligned}
\end{equation}
\end{widetext}

\begin{widetext}
\begin{equation}
\begin{aligned}
f&_{\rm{cal}}(E; \mu, \sigma, \lambda, \alpha) \\
 &=
   \alpha\DF{1}{\sigma \sqrt{2\pi}} e^{-\frac{(E-\mu) ^2}{2\sigma^{2}} }
  +(1-\alpha)\ \frac{\lambda e^\frac{\sigma^2\lambda^2+2\lambda E}{2}}{e^{\lambda\mu}-1} \left[
  \text{erf}\left(\frac{\mu-E-\sigma ^2\lambda}{\sqrt{2}\sigma}\right)-\text{erf}\left(\frac{-E-\sigma ^2\lambda}{\sqrt{2}\sigma}\right)\right]
  \rm{\ (exponential\ tail),}
\label{eq:exp}
\end{aligned}
\end{equation}
\end{widetext}
where both the peak and tail functions have been independently normalized to 1 over ($-\infty$, $+\infty$).

During the study a few more complicated tail shapes were also tested, such as combinations of constant and exponential tails:
\begin{equation}
\begin{aligned}
f_{\rm{cal}} &= [\alpha \delta+\beta f_{\rm{exp1}}+(1-\alpha-\beta)f_{\rm{exp2}}]\otimes \text{Gauss},
\label{eq:doubleExp}
\end{aligned}
\end{equation}
which contains two exponential tails, $f_{\rm{exp1}}$ and $f_{\rm{exp2}}$, with fractions of $\beta$ and $1-\alpha-\beta$, respectively, and
\begin{equation}
\begin{aligned}
\label{eq:ExpConst}
f_{\rm{cal}} &= [\alpha \delta+\beta f_{\rm{exp}}+(1-\alpha-\beta)f_{\rm{const}}]\otimes \text{Gauss},
\end{aligned}
\end{equation}
which contains one exponential tail, $f_{\rm{exp}}$, and one constant tail, $f_{\rm{const}}$, with fractions of $\beta$ and $1-\alpha-\beta$, respectively.

In addition, a simple case of energy dependent energy resolution can be found in Appendix~\ref{app:resolution}.

\section{Performance}
\label{sec:performance}
In this section we examine the performance of the calorimeter function.
Three practical cases were studied using \textsc{Geant4.10}:
uniformly distributed $\gamma$'s in a liquid scintillator detector,
a calibration source in the center of the scintillator detector,
and $\gamma$'s directed into a crystal EM calorimeter.
For each case, $1\times10^7$ events were simulated.
The simulation was configured to use the default electromagnetic model, and the minimal range cut was set to 1 mm.

All parameters in Eqs.~(\ref{eq:const}, \ref{eq:exp}, \ref{eq:doubleExp}, or \ref{eq:ExpConst}) were optimized.
The best-fit values of $\mu$, $\sigma$, and the number of totally absorbed events derived from peak fraction $\alpha$
were extracted and compared with the true values of:
\begin{enumerate}
\item \textbf{peak energy}, which is the total absorption energy,
\item \textbf{energy resolution}, which is the detector energy resolution, and
\item \textbf{peak area}, which is the number of totally absorbed events.
\end{enumerate}
The peak area is not explicitly defined for the CB function, so the area of its Gaussian (integrated from $-\infty$ to $+\infty$) was used as a proxy.
The $Accuracy$ is defined as the percentage error of the fit result with respect to the true value:
\begin{equation}
  Accuracy \equiv \left(\frac{\rm{fit\ result - true\ value}}{\rm{true\ value}}\right)\cdot 100\%.
\label{eq:accu}
\end{equation}

\subsection{Case 1 : Liquid scintillator detector}
\label{subsec:LS}
\begin{figure}[!ht]
	\begin{center}
		\includegraphics[width=0.4\columnwidth, clip]{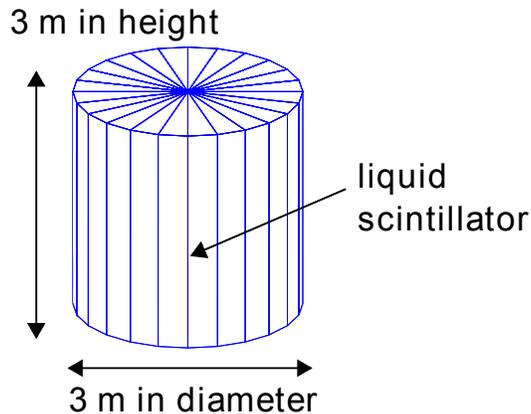}
	\end{center}
	\caption{(Color online) Liquid scintillator detector geometry.}
	\label{fig:ADgeo}
\end{figure}

Liquid scintillator (and similarly liquid argon, liquid neon, or liquid xenon) has been used by numerous neutrino and dark matter experiments, including the KamLAND~\cite{KLnH}, Double Chooz~\cite{DC}, RENO~\cite{RENO}, Daya Bay~\cite{DYB},
DEAP~\cite{DEAP}, and XMASS~\cite{XMASS} experiments.
It has also been selected for use by the forthcoming RENO50~\cite{RENO50}, LENA~\cite{LENA}, JUNO~\cite{JUNO}, SNO+~\cite{SNOplus}, and Jinping~\cite{Jinping} experiments. The detectors of these experiments are large electromagnetic calorimeters.
We simulated a cylindrical liquid scintillator detector of 3 meters in height and 3 meters in diameter, as shown in Fig.~\ref{fig:ADgeo},
which is the scale of the Double Chooz, RENO, and Daya Bay detectors.
The density of the liquid scintillator was 0.86~g/cm$^3$ with a carbon to hydrogen ratio of 17:28.
We simulated 2.2-MeV $\gamma$'s uniformly distributed in the detector.
The 2.2-MeV $\gamma$ from neutron capture by hydrogen has been used to measure neutrino oscillation and the reactor neutrino spectrum, in addition to being used for detector calibrations~\cite{KLnH, DCnH, nH, nH2}.

\begin{figure}[!htb]
	\begin{center}
		\includegraphics[width=0.47\columnwidth, clip]{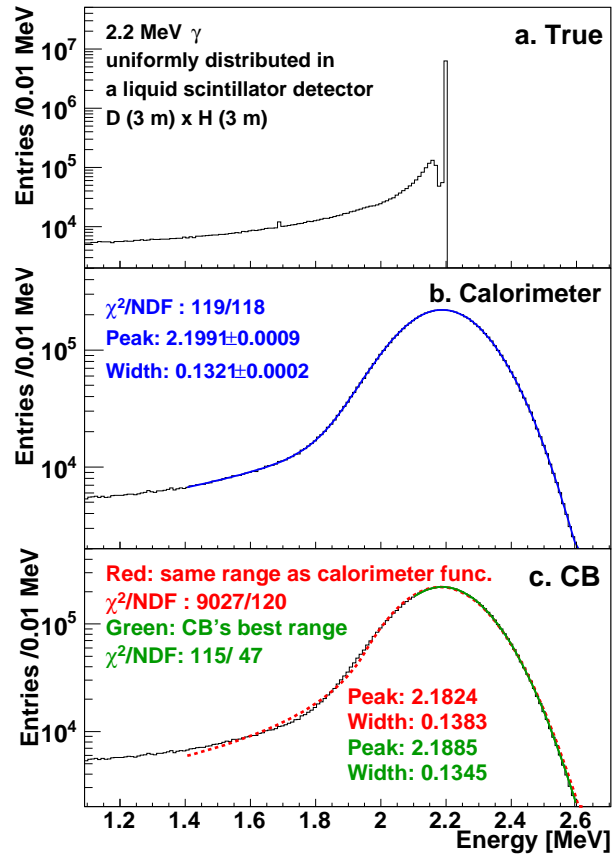}
	\end{center}
	\caption{(Color online) Plot (a) is the energy deposited in a liquid scintillator detector. Plot (b) shows the fitting result with the calorimeter function. Plot (c) shows the fitting results with the CB function with two different fitting ranges. The fit errors with the CB function are at the same level as with the calorimeter function.}
	\vspace{-0.5cm}
	\label{fig:ADfit}
\end{figure}

Each 2.2-MeV $\gamma$ developed an electromagnetic cascade in the detector.
For $\gamma$'s in the energy region of [0.1 MeV, 5 MeV] in the simulated liquid scintillator,
the primary interaction is Compton scattering, which comprises more than 90\% of the total scattering cross-section.
The other two minor processes are Rayleigh scattering and pair production.
The total absorption length is approximately 5 cm to 20 cm for photon energies from 0.1 MeV to 5 MeV~\cite{PDG}.
The true deposited energy spectrum of the 2.2-MeV $\gamma$ is shown in Fig.~\ref{fig:ADfit}(a).
The primary $\gamma$ and its secondary particles in the cascade sometimes escaped the scintillator resulting in partial energy absorption.
The Compton scattering edge of the 2.2-MeV $\gamma$ and those of the secondary $\gamma$'s pile up below 2.2 MeV.
A small peak at 1.69 MeV due to the escape of one annihilation photon is seen.
The $\approx$20-keV gap just below the peak is due to daughter $\gamma$'s that deposited most of their energy and then escaped the scintillator.
The reason the gap is about 20~keV is because that is the energy around which the interaction cross-section begins to increase exponentially (see, for example, Ref.~\cite{PDG, NIST}).
Thus a $\gamma$ with energy $<$ 20~keV is much less likely to escape the detector.

The deposited energy spectrum was smeared by a constant-resolution Gaussian function of $\sigma/\mu=6\%$,
which is similar to the Daya Bay experiment~\cite{DYB}.
This example made use of Eq.~(\ref{eq:doubleExp}), which contains two exponential tails.  In Fig.~\ref{fig:ADfit}(b),
an improvement in the $\chi^2/\rm{NDF}$ can be seen with respect to the CB function result, which is
fitted with the same range or its own best fit range, as shown in Fig.~\ref{fig:ADfit}(c).
The details are summarized in Table~\ref{tab:table1} for the $Accuracy$ of the peak energy, energy resolution, and peak area.
The relative statistical uncertainties for the peak, resolution, and peak area are 0.04\%, 0.15\%, and 2\%, respectively.
Compared with the fit results with the CB function, each $Accuracy$ is improved by an order of magnitude ($>$ 10).

\subsection{Case 2 : Calibration source}
Calibration sources are essential to make precise measurements using liquid scintillator detectors.
For example, $^{68}\rm{Ge}$ and $^{60}\rm{Co}$ sources have been used to calibrate the energy scale of a detector, or to study its non-linear energy response~\cite{ACU}. $^{68}\rm{Ge}$ decays through positron emission~\cite{nndc} where the positron quickly annihilates and emits two 0.511-MeV $\gamma$'s. $^{60}\rm{Co}$ undergoes $\beta$-decay and emits $\gamma$'s: over 99\% of the time, a 1.17-MeV and 1.33-MeV $\gamma$ will be observed~\cite{nndc}.

\begin{figure}[!ht]
	\begin{center}
		\includegraphics[width=0.47\textwidth, clip]{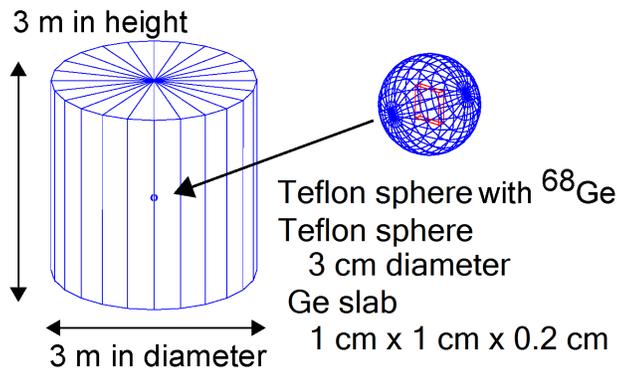}
	\end{center}
	\caption{(Color online) Calibration source in the center of a liquid scintillator detector.}
	\label{fig:Gegeo}
\end{figure}

A realistic $^{68}$Ge calibration source~\cite{ACU} was simulated in the center of the liquid scintillator detector modeled for the previous section.
The source geometry was simplified as a Teflon sphere of 3-cm diameter
with a thin $^{68}$Ge slab of 1~cm$\times$1~cm$\times$0.2~cm in its center.
The geometry of the setup is shown in Fig.~\ref{fig:Gegeo}.
The material of Teflon was simulated using Geant4 default settings with a density of 2.2~g/cm$^3$ and
a carbon to fluorine ratio of 2:4.

\begin{figure}[!h]
	\begin{center}
		\includegraphics[width=0.47\columnwidth, clip]{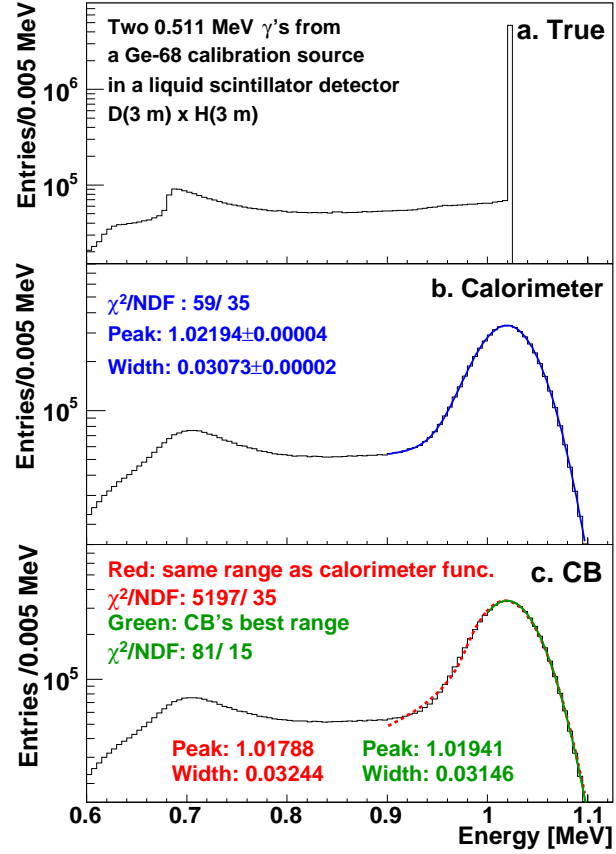}
	\end{center}
	\caption{(Color online) Plot (a) is the deposited energy of a $^{68}\rm{Ge}$ calibration source in a liquid scintillator detector. Plot (b) shows the fitting result with the calorimeter function. Plot (c) shows the fitting results with the CB function for two different fitting ranges. The fit errors with the CB function are at the same level as with the calorimeter function.}
\label{fig:Ge-new}
\end{figure}

Figure~\ref{fig:Ge-new}(a) shows the spectrum of energy deposited from the positron source.
Compton scattering is still the primary interaction process.
Undetectable energy loss occurred dominantly in the Teflon sphere but also beyond the region of liquid scintillator.
The resulting distribution is an overlap of two Compton scattering spectra.
One spectrum is the tail of energy deposited outside the liquid scintillator, which is similar to case 1, i.e.~Fig.~\ref{fig:ADfit}(a).
The other spectrum is due to the Compton scattering of one of the two 0.511-MeV $\gamma$'s in the Teflon.
The Compton edge of a 0.511-MeV $\gamma$ is at 0.34~MeV,
and the energy deposited in the liquid scintillator is 2$\times$0.511-0.34=0.68~MeV.
Since the $\gamma$'s were generated inside the Teflon sphere, the Compton edge is at the low-energy end,
which is opposite to the structure caused by energy deposition outside the liquid scintillator region.

A 3\% detector energy resolution smearing was applied, which is required by the JUNO experiment~\cite{JUNO}.
The fitting results of the smeared spectrum with the calorimeter function given by Eq.~(\ref{eq:exp}) and the CB function are shown in Fig.~\ref{fig:Ge-new}(b) and Fig.~\ref{fig:Ge-new}(c), respectively.
For the calorimeter function fitting, the relative statistical uncertainties for peak, resolution, and peak area are 0.004\%, 0.07\%, and 2\%, respectively.
An overall improvement of a factor of 10 can be found.
The comparison details are summarized in Table~\ref{tab:table1}.

\subsection{Case 3 : CsI crystal array}
\label{subsec:crystal}
\begin{figure}[!h]
	\begin{center}
		\includegraphics[width=0.47\columnwidth, clip]{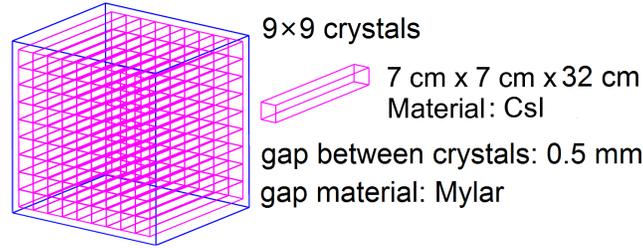}
	\end{center}
	\caption{(Color online) CsI crystal calorimeter geometry.}
\label{fig:EMCgeo}
\end{figure}

Last, we simulated an array of CsI crystals as used mostly in large scale particle detectors. Each CsI crystal was a cuboid of
7~cm$\times$7~cm$\times$32~cm, similar to the size of the BES III~\cite{BESIII} electromagnetic calorimeter.
7$\times$7 crystals were put together with a gap of 0.5~mm between adjacent crystals.
Each gap was filled with Mylar, which is often used as a wrapping material.
The geometry is shown in Fig.~\ref{fig:EMCgeo}.
50-MeV $\gamma$'s were directed toward the 
the central crystal.

\begin{figure}[!h]
	\begin{center}
		\includegraphics[width=0.47\columnwidth, clip]{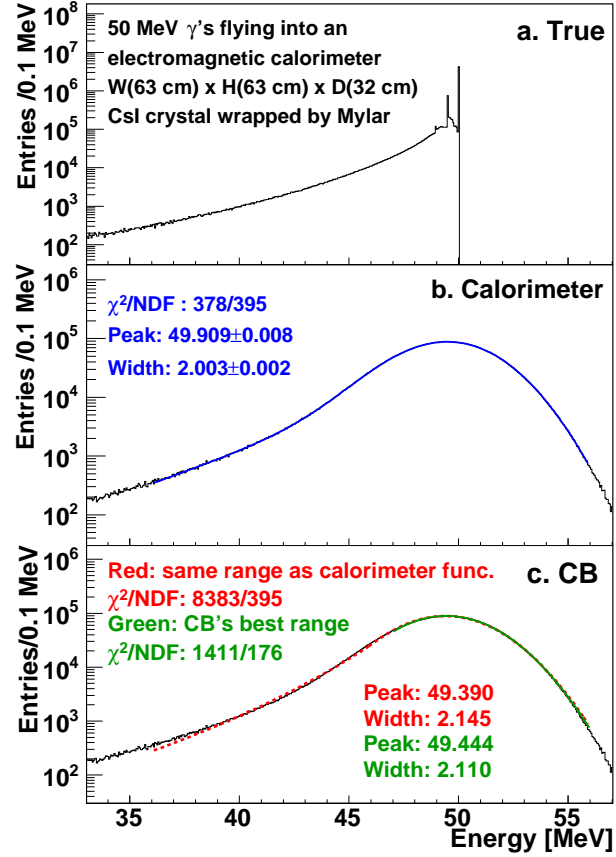}
	\end{center}
	\caption{(Color online) Plot (a) is the energy deposited in the CsI crystal calorimeter. Plot (b) shows the fitting result with the calorimeter function. Plot (c) shows the fitting results with the CB function with different fitting ranges. The fit errors with the CB function are at the same level as with the calorimeter function.}
\label{fig:EMC-new}
\end{figure}

Figure~\ref{fig:EMC-new} shows the distribution of deposited energy.
The dominant interaction process is pair production followed by Compton scattering.
Partial energy absorption occurred at every surface of the calorimeter and the Mylar gaps.  
The fraction of partial absorption is much large than for the previous two cases involving the liquid scintillator detector.
Below the total absorption peak is a peak due to the escape of a single annihilation $\gamma$.

The detector energy resolution was set to 4\% according to the BESIII~\cite{BESIII} detector.
The measured energy distribution was fitted with the calorimeter function given by Eq.~(\ref{eq:doubleExp}).
Because the tail has a complicated structure around the total absorption peak,
the two exponents of the tail function were predetermined by fitting the unsmeared tail shown in Fig.~\ref{fig:EMC-new}(a).
The fit results with the calorimeter and CB functions are shown in Fig.~\ref{fig:EMC-new}(b) and Fig.~\ref{fig:EMC-new}(c), respectively.
The relative statistical uncertainties of the fitting with calorimeter function for peak, resolution, and peak area are 0.02\%, 0.1\%, and 1\%, respectively.
Significant improvements in peak energy, energy resolution, and peak area were observed (see Table~\ref{tab:table1}).
Because the fraction of total absorption for this case is much smaller than for the previous two cases,
the extraction of the peak area is worse and would benefit from a more careful construction of the tail function.

\section{Discussion}
\label{sec:discussion}

A summary of the comparisons for all three cases is given in Table~\ref{tab:table1}.
The calorimeter function is successful in extracting the total absorption peak energy, energy resolution, and peak area,
with $Accuracy$ improved by an order of magnitude in almost all cases.

The CB function results in a lower peak energy and a larger energy resolution relative to the true values, in all cases.
This is due to the fact that the function does not consider the impact of the tail on the peak.
The peak area in the CB function is not a well-defined quantity, and was grossly overestimated.
The fourth column in Table~\ref{tab:table1} lists the results of the CB function using fit ranges that provided the best $\chi^2$/NDF.  These fits are essentially those of a truncated Gaussian, which can be seen from panel (c) of Figs.~\ref{fig:ADfit}, \ref{fig:Ge-new}, and \ref{fig:EMC-new}.

\begin{table}[!h]
		\begin{tabular}{ccccccc} \hline\hline
			                  &\multicolumn{2}{c}{Calorimeter}&\multicolumn{4}{c}{Crystal Ball}  \\
		    Fit range     &\multicolumn{2}{c}{Best fit}&\multicolumn{2}{c}{Same range}&\multicolumn{2}{c}{Best fit}\\
												                                                 \hline
			Liquid scintillator                                                                  	 \\
				peak accu.    &\multicolumn{2}{c}{-0.042\%}    &\multicolumn{2}{c}{-0.80\%}    &\multicolumn{2}{c}{-0.53\%}  \\
				resolution accu.   &\multicolumn{2}{c}{0.044\%}      &\multicolumn{2}{c}{4.8\%}     &\multicolumn{2}{c}{1.9\%}\\
				peak area accu.
				              &\multicolumn{2}{c}{-0.59\%}      &\multicolumn{2}{c}{ 22\% }     &\multicolumn{2}{c}{ 19\% }\\
				$\chi^2$/NDF  &\multicolumn{2}{c}{119/118}    &\multicolumn{2}{c}{9027/120}  &\multicolumn{2}{c}{115/47}\\
				                                                                               \\
			Calibration source                                                                 \\
				peak accu.    &\multicolumn{2}{c}{0.0054\%}    &\multicolumn{2}{c}{-0.40\%}    &\multicolumn{2}{c}{-0.25\%}\\
				resolution accu.   &\multicolumn{2}{c}{0.22\%}      &\multicolumn{2}{c}{5.8\%}     &\multicolumn{2}{c}{2.6\%}\\
				peak area accu.
				              &\multicolumn{2}{c}{-0.095\%}     &\multicolumn{2}{c}{18\%}     &\multicolumn{2}{c}{14\%}\\
				$\chi^2$/NDF  &\multicolumn{2}{c}{59/35}    &\multicolumn{2}{c}{5197/35}  &\multicolumn{2}{c}{81/15} \\
				                                                                               \\
		CsI crystal array                                                                     \\
				peak accu.    &\multicolumn{2}{c}{-0.18\%}    &\multicolumn{2}{c}{-1.2\%}    &\multicolumn{2}{c}{-1.1\%}\\
				resolution accu.   &\multicolumn{2}{c}{0.15\%}      &\multicolumn{2}{c}{7.3\%}    &\multicolumn{2}{c}{5.5\%}\\
				peak area accu.
				              &\multicolumn{2}{c}{17\%}         &\multicolumn{2}{c}{ 128\% } &\multicolumn{2}{c}{124\%}\\
				$\chi^2$/NDF  &\multicolumn{2}{c}{378/395}  &\multicolumn{2}{c}{8383/395} &\multicolumn{2}{c}{1411/176}\\

				\hline
		\end{tabular}
	\caption{\label{tab:table1}Summary for the performance of the calorimeter function and the CB function. The accu. ($Accuracy$) is defined in equation~(\ref{eq:accu}),
``peak'' means total absorption energy, ``resolution'' means energy resolution, and peak area is the number of totally absorbed events. The second column is for the three calorimeter functions.
The third column is for the CB function with the same range as each calorimeter function, while the fourth column is
with its own best fit range.
The sample statistics for each simulation is $1\times10^7$, and a coherent improvement can be found.}
\end{table}

The shape and fraction of the tail strongly depend on the structure of the detector as demonstrated in the previous sections.
Single or double exponential tail functions were generally sufficient for these applications.
However, additional forms can be explored by studying the energy tail using simulation,
which is the suggested approach when applying the calorimeter function.
As an example, the 20-keV gap of case 1 in Section~\ref{subsec:LS} may be better accommodated by integrating the tail function from 0 to $\mu - $20~keV (instead of from 0 to $\mu$).  A similar modification could be considered for the tail structure of case 3 in Section~\ref{subsec:crystal}.

At the same time, the shape and fraction of the tail are key to characterizing the detector structure and materials,
for example, to compare the amount of insensitive materials between data and simulation.
With the peak area and tail shape extracted, the calorimeter function can provide a unique perspective using data.


\section{Conclusion}
\label{sec:conclusion}
In this work, we studied energy deposition in an electromagnetic calorimeter.
Motivated by observed spectral features, an electromagnetic calorimeter function was developed.
The performance of the function was compared with the Crystal Ball function using simulation.
With a clear physical meaning, the accuracies of the extracted peak energy and energy resolution are significantly improved.
The concept of peak area is naturally defined.
The shape and fraction of the tail reflect the characteristics of the detector, and are thus informative
in understanding a detector.
A similar application to hadronic calorimeters and invariant mass distributions is expected.
Further studies are needed.

\section*{Acknowledgments}
ZW, LL, and SC are thankful for support from the
Key Laboratory of Particle \& Radiation Imaging, Ministry of Education,
Natural Science Foundation of China (No.~11235006 and No.~11475093),
and Ministry of Science and Technology of China (No.~2013CB834302).
JHC and GLL thank the
Ministry of Science and Technology, Taiwan under Grant No.~103-2112-M-009-018.

\appendix
\section{Including energy resolution}
\label{app:resolution}

The energy dependence of energy resolution is considered by performing the integration in Eq.~(\ref{eq:exp.tail}) under the assumption that the energy resolution is proportional to the square root of the deposited energy: $\sigma = c \sqrt{E'}$.

For an exponential tail and energy-dependent resolution, the measured energy distribution is
\begin{equation}
\begin{aligned}
f _{\rm{exponentail}} \otimes \text{Gauss} \ &= \int_{0}^{\mu} \lambda e^{\lambda E'}\cdot\DF{1}{c \sqrt{2\pi E'}}
e^{-\frac{(E'-E) ^2}{2c^{2}E'}}dE' \\
&= \frac{\lambda}{2A}e^{\frac{1-A}{c^2}E}
\left[
\text{erfc} \left( \frac{E-A\mu}{\sqrt{2c^2\mu}} \right) -
e^{\frac{2A}{c^2}E} \text{erfc} \left( \frac{E+A\mu}{\sqrt{2c^2\mu}} \right)
\right],
\label{eq:exp.tail.res}
\end{aligned}
\end{equation}
where $\mathrm{erfc}(E) \equiv 1 - \mathrm{erf}(E)$ and $A \equiv \sqrt{1-2c^2\lambda}$. Thus, this function is restricted to cases where the shape of the tail and the resolution satisfy $\lambda < \frac{1}{2c^2}$.

Combining Eqs.~(\ref{eq:derived}), (\ref{eq:delta peak}), and (\ref{eq:exp.tail.res}) after normalization, we obtain Eq.~(\ref{eq:res}) for the complete form of the function:

\begin{widetext}
\begin{equation}
\begin{aligned}
f&_{\text{cal}}(E; \mu, c, \lambda, \alpha)
=
\alpha\DF{1}{c \sqrt{2\pi E}} e^{-\frac{(E-\mu) ^2}{2c^2E} }
+(1-\alpha) \frac{\lambda}{2A} \frac{e^{\frac{1-A}{c^2}E}}{e^{\lambda\mu}}
\left[
\text{erfc} \left( \frac{E-A\mu}{\sqrt{2c^2\mu}} \right) -
e^{\frac{2A}{c^2}E} \text{erfc} \left( \frac{E+A\mu}{\sqrt{2c^2\mu}} \right)
\right].
\label{eq:res}
\end{aligned}
\end{equation}
\end{widetext}

\end{spacing}
\end{document}